# Controllable p-type Doping of 2D WSe$_2$ via Vanadium Substitution

Azimkhan Kozhakhmetov,[1] Samuel Stolz,[2] Anne Marie Z. Tan,[3,4] Rahul Pendurthi,[5] Saiphaneendra Bachu,[1] Furkan Turker,[1] Nasim Alem,[1,6,7] Jessica Kachian,[8] Saptarshi Das,[1,5,6] Richard G. Hennig,[4] Oliver Gröning,[2] Bruno Schuler,[2] Joshua A. Robinson.[1,7,9,*]

1. Department of Materials Science and Engineering, The Pennsylvania State University, University Park, PA 16802, USA
2. nanotech@surfaces Laboratory, Empa – Swiss Federal Laboratories for Materials Science and Technology, 8600 Dübendorf, Switzerland
3. School of Mechanical and Aerospace Engineering, Nanyang Technological University, Singapore, 639798, Singapore
4. Department of Materials Science and Engineering, University of Florida, Gainesville FL 32611, USA
5. Department of Engineering Science and Mechanics, The Pennsylvania State University, University Park, PA 16802, USA
6. Materials Research Institute, The Pennsylvania State University, University Park, PA 16802, USA
7. Two-Dimensional Crystal Consortium, The Pennsylvania State University, University Park, PA 16802, USA
8. Intel Corporation, 2200 Mission College Blvd, Santa Clara, CA 95054, USA
9. Center for 2-Dimensional and Layered Materials, The Pennsylvania State University, University Park, PA 16802, USA

* jrobinson@psu.edu

## Abstract

Scalable substitutional doping of two-dimensional (2D) transition metal dichalcogenides (TMDCs) is a prerequisite to developing next-generation logic and memory devices based on 2D materials. To date, doping efforts are still nascent. Here, we report scalable growth and vanadium (V) doping of 2D WSe$_2$ at front-end-of-line (FEOL) and back-end-of-line (BEOL) compatible temperatures of 800 °C and 400 °C, respectively. A combination of experimental and theoretical studies confirm that vanadium atoms substitutionally replace tungsten in WSe$_2$, which results in p-type doping via the introduction of discrete defect levels that lie close to the valence band maxima. The p-type nature of the V dopants is further verified by constructed field-effect transistors, where hole conduction becomes dominant with increasing vanadium concentration. Hence, our study presents a method to precisely control the density of intentionally introduced impurities, which is indispensable in the production of electronic-grade wafer-scale extrinsic 2D semiconductors.

**Introduction**

Doping transition metal dichalcogenides (TMDCs) is one of the major routes to realize next-generation logic transistors with ultimate gate length scaling. Both intrinsic and extrinsic TMDCs have been actively explored as a channel material at front-end-of-line (FEOL)[1–4] and as a diffusion barrier, liner, and thin-film transistor (TFT) at back-end-of-line (BEOL)[5–7] for advanced semiconductor technology nodes. To date, wafer-scale synthesis of near electronic-grade intrinsic TMDCs has been successfully realized at FEOL (>700 °C)[8–10] and BEOL(<500 °C)[11–14] compatible temperatures. However, while various doping techniques can tune the electrical conductivity of TMDCs (e.g. surface charge transfer doping,[15] electrostatic doping,[16] intercalation,[17] and substitutional doping,[18,19]) progress is still limited in truly scalable doping. Furthermore, even though "proof-of-concept" devices have been demonstrated based on doped TMDCs including vanadium dopants, [20–24] uniform distribution and precise control of the impurity density over a large scale and use of methods compatible with the state-of-the-art Si CMOS 300 mm process lines, still remains challenging.[19,25,26] Thus, scalable doping of TMDCs at a large temperature window with accurate control over the doping concentration is urgently needed.

Here, we report controlled substitutional vanadium doping of WSe$_2$ films realized via MOCVD at BEOL and FEOL compatible temperatures using W(CO)$_6$, H$_2$Se, and (C$_5$H$_5$)$_2$V as metal, chalcogen, and dopant sources, respectively. Accurate tuning of the partial pressures of the reactants allows the synthesis of V-doped WSe$_2$ with predetermined vanadium concentration, as verified by high-resolution x-ray photoelectron spectroscopy (XPS). Furthermore, room temperature Raman and photoluminescence (PL) spectroscopy studies reveal a strong dopant concentration dependence of the spectra where defect activated modes increase in intensity and positive trions are formed that completely quench the PL spectra. Atomic-scale characterization techniques such as aberration corrected scanning transmission electron microscopy (STEM), scanning tunneling microscopy and spectroscopy (STM/STS), and density functional theory (DFT) calculations reveal that vanadium atoms substitutionally replace W atoms in the WSe$_2$ lattice and introduce multiple defect states that are in close proximity to the valence band edge. The p-type nature of the vanadium dopants is further confirmed with back-gated field-effect transistors that exhibit an enhanced p-branch (hole) current and a positive threshold shift. Our systematic study provides substantial progress and fundamental insight in large-scale synthesis of extrinsic 2D TMDCs with precisely controlled dopant concentrations that will be a major step towards successful integration with Si CMOS at FEOL and BEOL.

**Results and Discussion:**

Pristine and V-doped FEOL compatible WSe$_2$ is synthesized on c-plane sapphire substrates at 800 °C and 700 Torr using W(CO)$_6$, V(C$_5$H$_5$)$_2$ and H$_2$Se, respectively.[8,18] The introduction of vanadium atoms has a direct impact on the surface morphology of the obtained films, which are fully coalesced and primarily monolayer with some secondary-islands on top. (Figure 1a, b). The domain size is 800-900 nm in extrinsic films, where the introduction of vanadium dopants leads to a deviation from the traditional triangular shape for intrinsic WSe$_2$, becoming truncated triangles and hexagons (Figure S1a-d). Furthermore, an increase in density of bilayer islands occurs in extrinsic films with increasing dopant concentration (Figure S1c, d), suggesting that vanadium adatoms serve as secondary nucleation centers on the surface.[18] In parallel, BEOL-compatible WSe$_2$ is obtained on SiO$_2$/Si substrates at 400 °C and 700 Torr, with all films being coalesced, multilayer, and polycrystalline with an average domain size of 100 nm and an average thickness of 2.11 nm, corresponding to 3-4 layers (Figure S2a-d and S3a, b).[11] Moreover, chemical composition analysis via XPS indicates that V is incorporated into the WSe$_2$ lattice at both FEOL and BEOL temperatures.[27] Both intrinsic and extrinsic FEOL films exhibit characteristic W 4f$_{7/2}$, W 4f$_{5/2}$, and W 5p$_{3/2}$ peaks at 32.6 eV, 34.8 eV, 38.1 eV and Se 3d$_{5/2}$ and Se 3d$_{3/2}$ peaks at 54.8 eV and 55.7 eV, respectively (Figure 1c and S4a) agreeing with previous reports.[18,28] As the vanadium concentration increases, associated V 2p$_{3/2}$ and V 2p$_{1/2}$ XPS peaks are detectable at 513.7 eV and 521.7 eV, respectively, corresponding to V-Se bonding and indicating successful doping of the WSe$_2$ with V atoms (Figure 1d and S3b).[29] Moreover, a peak at ~517.0 eV is observed that corresponds to V-O bonding, and suggests vulnerability of extrinsic films towards oxidation (Figure 1d and Figure S4b).[30,31] We observe similar characteristic W 4f, Se 3d and V 2p peaks in XPS on BEOL V-WSe$_2$ on SiO$_2$/Si, where the dopant concentration varies as a function of the vanadium precursor flow rate (Figure S5a, b).

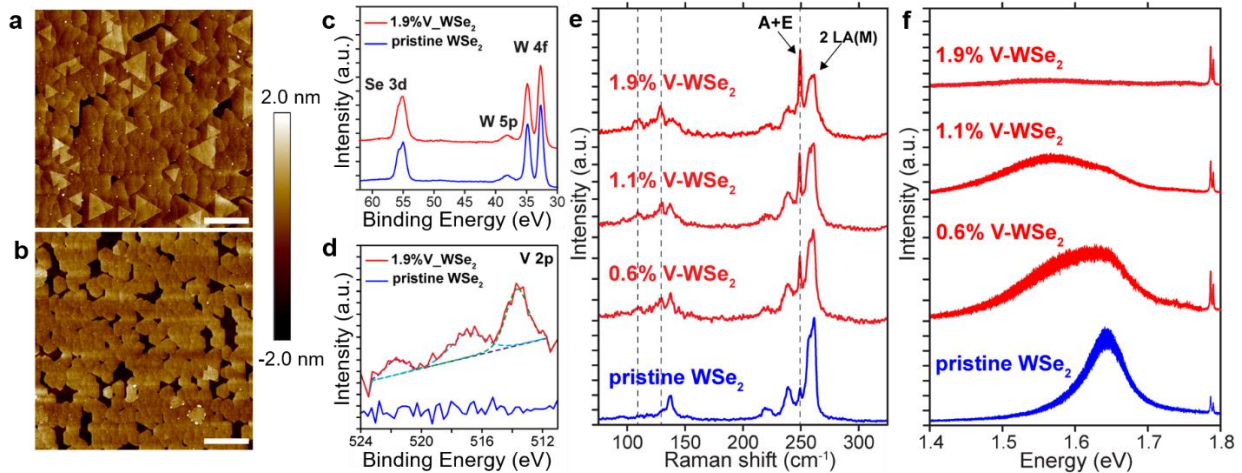

**Figure 1.** Surface morphology, chemical composition, and optical properties of pristine and V-doped FEOL WSe$_2$ films on sapphire substrates. AFM images of (a) pristine and (b) 1.9% vanadium doped WSe$_2$ films demonstrating a clear impact of the dopants on the surface morphology (1μm scale bar). (c, d) High-resolution XPS analysis further confirms the presence of vanadium atoms in the lattice where V 2p spectra are found at 513.7 eV, 517.0 eV, and 521.7 eV. (e, f) Corresponding Raman spectra display increased density of defect activated modes whereas PL spectra start to redshift and are completely quenched (1.9 % V-WSe$_2$) as a function of vanadium concentration, respectively.

Vanadium doping leads to enhancement in Raman defect modes and quenching of WSe$_2$ photoluminescence. Intrinsic and extrinsic FEOL compatible films exhibit peaks at 249.3 cm$^{-1}$ and 260.6 cm$^{-1}$ corresponding to A+E and 2 LA(M) characteristic in-plane and out-of- plane Raman active modes of 2D WSe$_2$.[32,33] Intensified defect activated modes (ZA(M) (109.2 cm$^{-1}$) and LA(M) (128.6 cm$^{-1}$)) are the direct result of the increased disorder in the WSe$_2$ lattice due to the introduction of vanadium and their intensity can serve as measure for the dopant concentration.[34,35] The corresponding PL spectra exhibit a similar dopant concentration dependence where intrinsic WSe$_2$ displays a 1.65 eV optical band gap , while V-WSe$_2$ is redshifted by 40 meV (0.6% V-WSe$_2$) and 80 meV (1.1% V-WSe$_2$) with reduced PL intensity. As the dopant concentration increases to 1.9%, there is a complete quench in the PL. The observed PL quenching is hypothesized to be due to an increased density of positive trions because vanadium is expected to be a p-type dopant in WSe$_2$, and donates an extra hole to the system.[25,36,37] Charged trions, unlike neutral excitons experience stronger electrostatic interaction with the dopant impurities resulting in non-radiative recombination that quenches the PL intensity for vanadium concentrations >1.9%.[16,38] As expected, BEOL V-WSe$_2$ films display similar Raman and PL properties (Figure S5c, d) as a function of vanadium concentration where the host lattice is perturbed after the dopant incorporation.

STM reveals a variety of uniformly distributed point defects in V-doped WSe$_2$ (Figure S6). Several types of point defects were previously identified as either Re (~32 ppm), Mo (~340 ppm), or Cr (~90 ppm) substituting for W,[18,39] or as oxygen substituting for selenium (~19 ppm),[40] respectively. Consequently, we assign the most common point defect (0.43%), which appears as a dark depression at positive sample bias (Figure 2a) in the STM topography, as vanadium impurities. At negative sample bias, however, vanadium dopants appear as a bright protrusion (Figure 2b), indicative of a negatively charged (ionized) impurity.[41] Indeed, Kelvin probe force microscopy (KPFM) experiments (Figure 2g) corroborate V impurities to be negatively charged by the positive shift of the local contact potential difference (LCPD).[42]

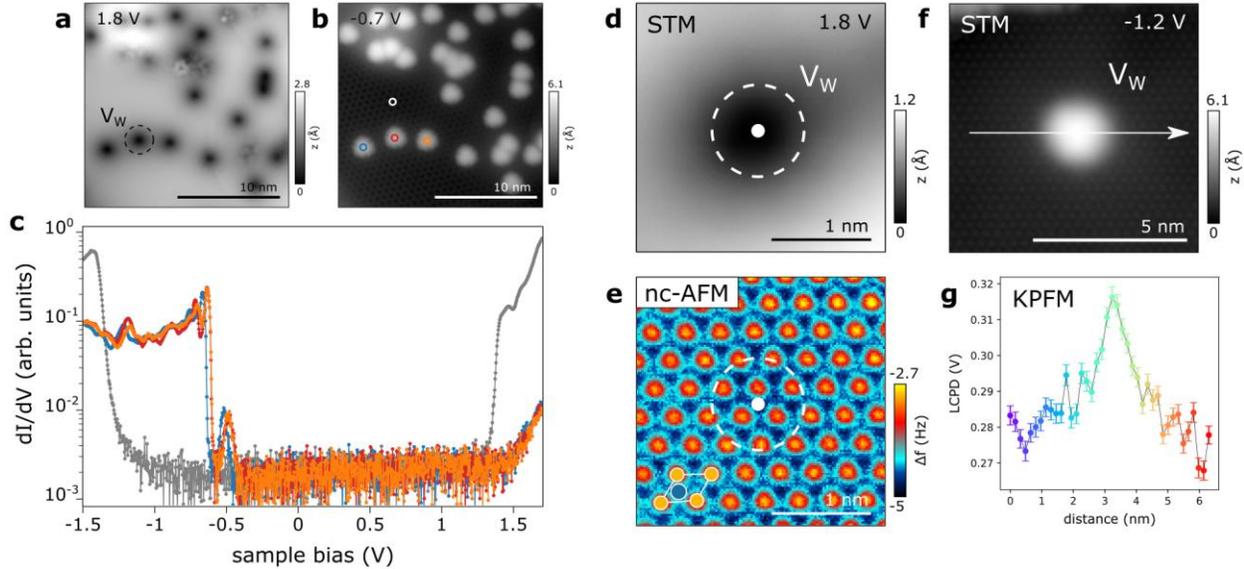

**Figure 2.** (a,b) STM images of V-doped WSe$_2$ on EG/SiC at the same location at positive (V = 1.8 V) and negative (V = -0.7 V) sample bias. Multiple $V_W$ dopants at various separations can be recognized. (c) Differential conductance (dI/dV) spectra measured on three $V_W$ dopants in color (locations indicated by circles in b) and the bare WSe$_2$ in gray. The spectra feature a series of p-type defect states above the valence band edge. (d,e) STM and CO-tip nc-AFM image of a single V substituent at a W site. The $V_W$ position is marked by a white dot. The unit cell is indicated at the bottom left in (e) (W = blue, Se = orange). (f) STM image of $V_W$ at negative (V = -1.2 V) sample bias. (g) Kelvin probe force microscopy (KPFM) measurement across the $V_W$ defect (along arrow in f) exhibiting a positive shift of the local contact potential difference (LCPD), indicative of a negative defect charge.

Combining STM and CO-tip noncontact atomic force microscopy (nc-AFM) confirms that V impurities substitutionally replace W atoms in the WSe$_2$ lattice (Figure 2d,e). The STS spectra of individual, relatively isolated V dopants (Figure 2c) reveal multiple in-gap states in the occupied spectrum, close to the valence band edge, in agreement with the expected formation of p-type defect states. At positive sample bias, the conduction band onset is pushed towards higher energies in the vicinity of dopants resulting from local band bending.[43] When the dopants are very close to each other, the defects exhibit a spectrum that is shifted towards higher energies (Figure S7c), an electrostatic effect due to the negative defect charge analog to the negative energy shift observed for positively charged n-type dopants.[18] This effect can be as large as several hundred millivolts, shifting the most tightly bound defect state up to the Fermi level.

Atomic resolution high angle annular dark field-scanning transmission electron microscopy (HAADF – STEM) confirms the presence of vanadium atoms in the lattice (Figure 3a). The technique relies on atomic number based contrast difference[44] where vanadium atoms are expected to exhibit reduced intensity as compared to W atoms owing to their smaller atomic number. The dopants (green dashed circles) are uniformly distributed throughout the WSe$_2$ lattice. The extrinsic films also contain other types of point defects such as single Se vacancies (red dashed circles) and double Se vacancies (cyan dashed circles) that are commonly observed in the WSe$_2$ lattice[45] (Figure 3a) and attributed in part to beam-induced knock-on damage by the electron beam.[46] Moreover, selected area electron diffraction (SAED) pattern exhibits single crystal signature with hexagonal symmetry (Figure 3a inset) indicating crystalline structure of V-WSe$_2$ films. An intensity line profile obtained along an armchair direction containing one V-dopant confirms the substitutional nature of the V dopant atom at the W site of the lattice (Figure 3b and inset) supporting our previous observations via XPS, Raman and PL. This is evident from the low intensity peak corresponding to the V dopant atom whereas an adatom results in a higher intensity peak due to the combined intensity from W and V atoms. Furthermore, STEM-energy dispersive spectroscopy (EDS) mapping (Figure S8a-e) underlines the uniform distribution of constituent elements in the film.

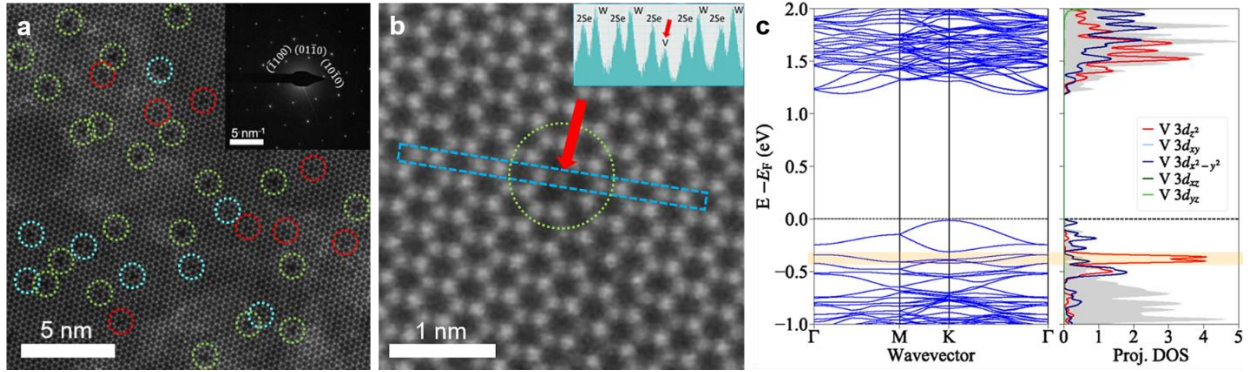

**Figure 3.** Atomic resolution imaging and density functional theory (DFT) calculation of band structure, projected density of states and the possible position of vanadium dopants in the WSe$_2$ lattice. (a) HAADF-STEM image showing V dopants (green dashed circles), single Se vacancies (red dashed circles) and double Se vacancies (cyan dashed circles) and SEAD pattern (inset) displaying single crystal nature of the films with hexagonal symmetry. (b) Higher magnification HAADF-STEM image showing one V dopant atom where the blue box indicates an armchair direction along which the intensity profile is obtained (inset) highlighting the substitutional doping of the V dopant atom. (c) Calculated band structure and projected density of states for the ionized V dopant on a W site, including spin–orbit coupling. The density of states is projected onto the $d$-orbitals of the V dopant atom, with the gray shading indicating the density of states for pristine WSe$_2$, for comparison. The most relevant defect state introduced by the V dopant near the top of the valence band is highlighted in the orange box.

First-principles DFT calculations (see Materials and Methods) confirm that the W site is the energetically most favorable site in the WSe$_2$ lattice for the V dopant (Table S1), agreeing with the experimental findings. No significant atomic relaxations or symmetry breaking of the lattice in the vicinity of the dopant is observed in the calculations. The DFT-computed electronic band structure and density of states (Figure 3c) indicate that the V dopant introduces energy states (highlighted in orange) close to the top of the valence band. By projecting the density of states onto the individual $d$-orbitals of the V dopant atom, this energy state is primarily from the V $3d_z^2$ orbital. When in the neutral (un-ionized) state, the dopant level is only partially occupied; however, the dopant is easily ionized, becoming negatively charged, and the fully filled dopant level falls slightly below the top of the valence band (Figure 3c) agreeing well with our experimental data and previous reports.[37]

Transport measurements further corroborate the p-type nature of vanadium dopants. Back-gated field effect transistors (BGFETs) were fabricated on an Al$_2$O$_3$/Pt/TiN/p$^{++}$-Si substrate using standard electron beam lithography process described elsewhere.[47,48] Transfer characteristics, i.e., source to drain current ($I_{DS}$) as a function of back-gate voltage ($V_{BG}$) for different source to drain voltage ($V_{DS}$) of pristine and lightly doped (0.6%) WSe$_2$ show ambipolar characteristics (Figure 4 a, b), with both n- and p-branches present. However, increasing the doping concentration to 1.1% and 1.9% (Figure 4 c ,d) leads to enhancement in the p-branch and reduction of the n-branch within the same $V_{BG}$ measurement window. In other words, positive shift in the transfer characteristics or threshold voltage ($V_{th}$) of the p-branch with increasing V concentrations confirms the p-dopant nature of the V atoms in the WSe$_2$, deduced from STM experiments and DFT simulations.

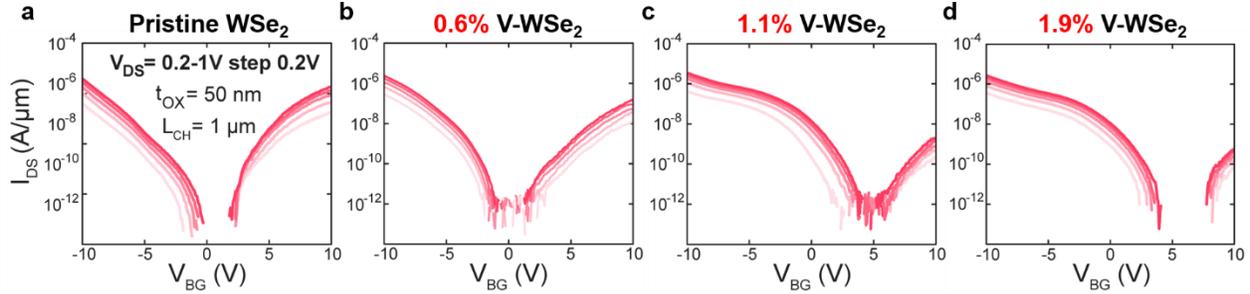

**Figure 4.** Transport properties of intrinsic and extrinsic FEOL WSe$_2$ back-gate field-effect transistors (BGFETs) transferred on 50 nm ALD Al$_2$O$_3$ substrates. (a) Transfer characteristics of the intrinsic films ($I_{DS}$ vs $V_{BG}$) at a different drain voltages ($V_{DS}$) display expected ambipolar conduction. (b)-(d) However, electron current (n-branch) starts to degrade as a function of vanadium concentration and BGFETs are no longer ambipolar and predominantly p-type.

The hole mobility is extracted in the linear regime of the transfer curve using the peak transconductance method given by the equation:

$$\mu_h = \frac{dI_{DS}}{dV_{BG}} \left( \frac{W}{L_{CH}} C_{ox} V_{DS} \right)^{-1}$$

where $W$ is the width, $L_{CH}$ is the length of the channel, $C_{ox}$ is the gate capacitance per unit area given as $C_{ox} = \frac{\varepsilon_0 \varepsilon_{ox}}{t_{ox}}$, where $\varepsilon_0$ is the dielectric constant of free space, $\varepsilon_{ox}$ is the dielectric constant of Al$_2$O$_3$, and $t_{ox}$ is the thickness of the gate dielectric. The threshold voltage ($V_{th}$) was extracted using the constant current method for $I_{DS}$ = 100 nA/µm (Table S2). All the constructed devices (Table S2) display ON-OFF ratio ($I_{ON}/I_{OFF}$) exceeding $10^6$ and mobility values similar to previous reports.[25,49] Improvement in the $I_{ON}$ for the p channel, without losing gate modulation, further highlights the effectiveness of vanadium for p-doping of large-area 2D WSe$_2$ at FEOL and BEOL compatible temperatures.

**Conclusion**

The scalable synthesis and vanadium doping of 2D WSe$_2$ is readily achievable via MOCVD at FEOL and BEOL compatible temperatures. Our experimental findings confirm the successful incorporation of the dopants by substitutionally replacing W atoms in the host lattice where the concentration of vanadium atoms can be precisely tuned by having accurate control over the vanadium precursor flux. Atomic-scale characterization reveals that the dopants are uniformly distributed throughout the WSe$_2$ lattice. Moreover, incorporated V atoms introduce discrete energy levels within the band gap of WSe$_2$, located in proximity to its valence band and thus confirm the expected p-type nature of the dopants. Ultimately, the integration of 2D materials with Si CMOS will require wafer-scale production of intrinsic and extrinsic materials at various temperatures with the highest precision. In this context, the method introduced in the current study will help to bridge the gap between academia and industry.

**Experimental Section:**

*MOCVD growth of pristine and V-doped WSe$_2$ films.* A custom-designed vertical cold wall gas-source CVD reactor is utilized to synthesize intrinsic and extrinsic WSe$_2$ films as previously reported.[11,18] The tungsten hexacarbonyl (W(CO)$_6$) (99.99 %, Sigma-Aldrich), Bis(cyclopentadienyl)vanadium (V(C$_5$H$_5$)$_2$) (sublimed, 95%, Strem Chemicals), and hydrogen selenide (H$_2$Se) (99.99% Matheson) are used as metal, dopant and chalcogen precursors, respectively in a 100% H$_2$ ambient. The solid-state metal and dopant sources are kept inside the stainless-steel bubblers where temperature and pressure of the bubblers are constantly maintained at 25 °C and 725 Torr, and 40 °C and 725 Torr, respectively. Hydrogen selenide is supplied from a different gas manifold and all three precursors are introduced from separate lines to prevent the intermixing before reaching the reactor inlet. The three-step growth (nucleation, ripening, and lateral growth) method[8] is used to grow monolayer, epitaxial intrinsic, and extrinsic FEOL WSe$_2$ on c-plane sapphire (Cryscore Optoelectronic Ltd, 99.996%) substrates. Growth temperature, pressure, and H$_2$Se flow rate are kept constantly at 800 °C, 700 Torr, and 7 sscm, respectively, for all three growth steps. At the nucleation stage, the metal and the dopant precursors are introduced simultaneously with a flow rate of 20 sccm and 60 sccm, respectively, for 2 min. At the ripening stage, W(CO)$_6$ and V$_2$(C$_5$H$_5$) were switched off and formed V-WSe$_2$ nuclei are let anneal under H$_2$Se for 10 min. W(CO)$_6$ is re-introduced with a constant 4.5 sccm at the lateral growth stage (30 min), whereas the dopant flow rate varied from 30 sccm to 5 sccm depending on the desired vanadium content in the films. After the ripening stage, the metal and dopant sources are cut-off and post-growth annealing under H$_2$Se is carried out for 10 min and followed by colling down to the room temperature. The BEOL counterparts are synthesized with a single-step growth method on SiO$_2$/Si substrates (University Wafer, Silicon P/B (100)) with 300 nm wet thermal oxide where the temperature and pressure of the reaction chamber are kept at 400 °C and 700 Torr, respectively.[11,18] To promote the lateral growth of the pristine and V-WSe$_2$ films at this kinetically limited growth regime, the precursors flow rates were adjusted accordingly to 2 sccm and 9 sccm for W(CO)$_6$ and H$_2$Se, respectively whereas the V$_2$(C$_5$H$_5$) flow rate is varied from 10 to 2.5 sccm. At both FEOL and BEOL conditions, the dopant concentration is a direct function of the (V(C$_5$H$_5$)$_2$) partial pressure where increasing the flow rate of the vanadium precursor results in a monotonic increase on the dopant concentration in the films. Within the investigated window of vanadium dopant concentrations, no saturation of the dopant concentration is observed at both FEOL and BEOL growth temperatures. Prior to the growth, sapphire and SiO$_2$/Si substrates are cleaned with acetone and isopropyl alcohol (IPA) in an ultrasonication bath for 10 min each followed by DI water rinse and N$_2$ gun dry. To further minimize the organic contamination on the surface, the substrates are cleaned with commercially available heated Piranha solution (Nanostrip, KMG Electronic Chemicals) at 90 °C for 20 min and rinsed with deionized (DI) water.

*Atomic Force Microscopy (AFM).* Bruker Icon I tool is used to acquire AFM data in a peak force tapping mode.

*Raman and Photoluminescence (PL) spectroscopy.* Raman and PL spectra of the samples are obtained by the Horiba Labram HR Evolution VIS-NIR Raman system with a 633 nm laser at 0.4 mW power and 532 nm laser at 0.4 mW power, respectively.

*Scanning Electron Microcopy (SEM).* Verios G4 with the accelerating voltage of 2 keV is employed to analyze the films.

*X-ray Photoelectron Spectroscopy (XPS).* High-resolution X-ray photoelectron spectroscopy data is obtained by a Physical Electronics Versa Probe II tool with a monochromatic Al K$_\alpha$ x-ray source (hv=1486.7 eV) at high vacuum (<10$^{-6}$ Torr) environment. The acquired spectra are charge corrected to C1s core level at 284.8 eV and W 4f$_{7/2}$ at 32.7 eV, respectively. U 2 Tougaard background is used for V and W whereas Iterated Shirley background is used for Se to fit the XPS spectra of the samples. The concentration of the vanadium dopants is quantified via the Casa XPS software where the peak area and intensity of the V 2p region are taken to account relative to W and Se regions.

*STM, STS, nc-AFM and KPFM.* V-doped WSe$_2$ FEOL samples were prepared *ex situ* on EG on SiC substrates followed by a final 450 °C anneal in ultrahigh vacuum. The measurements were performed with a commercial QPlus from Scienta Omicron operated at 5 K and at pressures below 2×10$^{-10}$ mbar. For STM, STS and KPFM the tungsten tip mounted on a QPlus tuning fork sensor was prepared on a clean Au(111) surface (sputtering: 10 min, Ar+, 1kV; annealing: 10 min, 450 °C) and confirmed to be metallic. For nc-AFM experiments, the metallic tip was modified with a single CO molecule[50] that was picked up from a Au(111) surface. STM topographic measurements were taken in constant current mode with the bias voltage given with respect to the sample. STS measurements were recorded using a lock-in amplifier at 610 Hz and a modulation amplitude of 20 meV. The nc-AFM topographies and KPFM measurements were acquired in constant height mode, while the QPlus sensor was driven at its resonance frequency (of about 25 kHz) with a constant amplitude of 70 pm. The frequency shift from resonance of the tuning fork was recorded using Omicron Matrix electronics and HF2Li PLL from Zurich Instruments.

*Transmission Electron Microscopy (TEM).* As grown V-WSe$_2$ films are lifted from sapphire substrates and transferred to Quantifoil Cu TEM grids using a PMMA assisted transfer method.[51] Selected area electron diffraction (SAED) patterns and scanning transmission electron microscopy-energy dispersive spectroscopy (STEM-EDS) maps are acquired using a Thermo Fisher Talos F200X instrument at 80 kV acceleration voltage. A Thermo Fisher Tian$^3$ G2 microscope equipped with image and probe correctors is employed to collect atomic resolution high angle annular dark field (HAADF)-STEM data. The images are acquired using 30 mrad semi convergence angle, ~ 50 pA screen current and 80 kV acceleration voltage to minimize the possible damage to the samples. Acquired atomic resolution images are smoothed with a 2-pixel gaussian blur filter using ImageJ where brightness and contrast of the images are adjusted accordingly.

*DFT calculations.* The formation energy and electronic structure of the substitutional V dopant was computed using DFT as implemented in the Vienna ab initio simulation package VASP.[52] Calculations were performed using projector-augmented wave potentials[53,54] with Perdew–Burke–Ernzerhof (PBE)[55] generalized gradient approximation functionals to treat the exchange-correlation. Spin-polarized calculations with spin-orbit coupling were performed with a plane wave cutoff energy of 520 eV, Methfessel–Paxton smearing[56] with a smearing energy width of 0.10 eV, and Γ-centered Monkhorst-Pack k-point meshes[57] for Brillouin zone integration. The dopant was modeled using 4 × 4 and 5 × 5 supercells containing a single dopant atom each, and 20 Å vacuum spacing between layers to minimize interlayer interaction.

The formation energy $E^f[X]$ of a neutral point defect $X$ was determined from DFT calculations using a supercell approach following $E^f[X] = E_{tot}[X] - E_{tot}[\text{pristine}] - \Sigma_i n_i \mu_i$, where $E_{tot}[X]$ and $E_{tot}[\text{pristine}]$ are the total DFT-computed energies of the supercell containing the defect $X$ and the pristine supercell respectively, $n_i$ is the number of atoms of species $i$ added/removed, and $\mu_i$ is the corresponding chemical potential of the species. The values reported in this work were referenced to bcc V for $\mu_V$, and either bcc W for $\mu_W$ (W-rich limit), or hexagonal (gray) Se for $\mu_{Se}$ (Se-rich limit).

*Fabrication of field effect transistors.* The as-grown pristine and V-doped WSe$_2$ films are transferred onto a 50 nm ALD Al$_2$O$_3$ on Pt/TiN/p++ Si from the host c-plane sapphire substrates. Photoresist is spun and 1μm by 5μm strips of V-WSe$_2$ are isolated from the conformal film using e-beam lithography (EBPG 5200 Vistec). After developing, a Plasma-Therm Versalock 700 inductively coupled plasma etch tool is used ($SF_6/O_2$ 30/10 sccm gas mixture) to remove the exposed film. Photoresist is spun and source and drain contacts are defined using e-beam lithography, followed by 40 nm Ni and 30 nm Au contact metal deposition in a Temescal e-beam evaporator tool. The electrical characterization is conducted using a Keysight B1500A semiconductor parameter analyzer; the measurements are done at room temperature and at ~$10^{-5}$ Torr.

**Acknowledgements**

A.K. and J.A.R. acknowledge Intel through the Semiconductor Research Corporation (SRC) Task 2746, the Penn State 2D Crystal Consortium (2DCC)-Materials Innovation Platform (2DCC-MIP) under NSF cooperative agreement DMR- 1539916, and NSF CAREER Award 1453924 for financial support. S.S. acknowledges funding from the Swiss National Science Foundation under SNSF project number 159690. A.M.Z.T. and R.G.H. were also funded by the NSF through the 2DCC-MIP under award DMR-1539916, and by additional awards DMR-1748464 and OAC-1740251. Computational resources were provided by the University of Florida Research Computing Center. The work of R.P. and S.D was supported by Army Research Office (ARO) through Contract Number W911NF1920338. S.B. and N.A. also acknowledge additional support provided by NSF CAREER DMR-1654107 and Materials Characterization Lab at Penn State.

**Conflict of Interest**

Authors declare no conflict of interest.

**References:**

[1] Y. Liu, X. Duan, H. Shin, S. Park, Y. Huang, X. Duan, *Nature* **2021**, *591*, 43.

[2] D. Akinwande, C. Huyghebaert, C. H. Wang, M. I. Serna, S. Goossens, L. J. Li, H. S. P. Wong, F. H. L. Koppens, *Nature* **2019**, *573*, 507.

[3] I. Asselberghs, Q. Smets, T. Schram, B. Groven, D. Verreck, G. Afzalian, G. Arutchelvan, A. Gaur, D. Cott, T. Maurice, S. Brems, K. Kennes, A. Phommahaxay, E. Dupuy, D. Radisic, J. De Marneffe, A. Thiam, W. Li, K. Devriendt, C. Huyghebaert, D. Lin, M. Caymax, P. Morin, I. P. Radu, *2020 IEEE Int. Electron Devices Meet.* **2020**, 40.2.1.

[4] Z. Ahmed, A. Afzalian, T. Schram, D. Jang, D. Verreck, Q. Smets, P. Schuddinck, B. Chehab, S. Sutar, G. Arutchelvan, A. Soussou, I. Asselberghs, A. Spessot, I. P. Radu, B. Parvais, J. Ryckaert, M. H. Na, *2020 IEEE Int. Electron Devices Meet.* **2020**, 22.5.1.

[5] C. Lo, B. A. Helfrecht, Y. He, D. M. Guzman, N. Onofrio, S. Zhang, *J. Appl. Phys* **2020**, *128*.

[6] J. A. Robinson, *APL Mater.* **2018**, *6*.

[7] A. Kozhakhmetov, R. Torsi, C. Y. Chen, J. A. Robinson, *J. Phys. Mater* **2020**, *4*.

[8] X. Zhang, T. H. Choudhury, M. Chubarov, Y. Xiang, B. Jariwala, F. Zhang, N. Alem, G. C. Wang, J. A. Robinson, J. M. Redwing, *Nano Lett.* **2018**, *18*, 1049.

[9] M. Seol, M. H. Lee, H. Kim, K. W. Shin, Y. Cho, I. Jeon, M. Jeong, H. I. Lee, J. Park, H. J. Shin, *Adv. Mater.* **2020**, *32*, 1.

[10] M. Chubarov, T. H. Choudhury, D. R. Hickey, S. Bachu, T. Zhang, A. Sebastian, A. Bansal, H. Zhu, N. Trainor, S. Das, M. Terrones, N. Alem, J. M. Redwing, *ACS Nano* **2021**, *15*, 2532.

[11]   A. Kozhakhmetov, J. R. Nasr, F. Zhang, K. Xu, N. C. Briggs, R. Addou, R. Wallace, S. K. Fullerton-Shirey, M. Terrones, S. Das, J. A. Robinson, *2D Mater.* **2020**, *7*.

[12]   H. Medina, J.-G. Li, T.-Y. Su, Y.-W. Lan, S-H. Lee, C.-W. Chen, Y.-Z. Chen, A. Manikandan, S.-H. Tsai, A. Navabi, X. Zhu, Y.-C. Shih, W.-S. Lin, J.-H. Yang, S. R. Thomas, B.-W. Wu, C.-H. Shen, J.-M. Shieh, H.-N. Lin, A. Javey, K. L. Wang, Y.-L. Chueh, *Chem. Mater.* **2017**, *29*, 1587.

[13]   J. Mun, Y. Kim, I. S. Kang, S. K. Lim, S. J. Lee, J. W. Kim, H. M. Park, T. Kim, S. W. Kang, *Sci. Rep.* **2016**, *6*, 1.

[14]   A. Kozhakhmetov, T. H. Choudhury, Z. Y. Al Balushi, M. Chubarov, J. M. Redwing, *J. Cryst. Growth* **2018**, *486*, 137.

[15]   A. J. Arnold, D. S. Schulman, S. Das, *ACS Nano* **2020**, *14*, 13557.

[16]   D. H. Lien, S. Z. Uddin, M. Yeh, M. Amani, H. Kim, J. W. Ager, E. Yablonovitch, A. Javey, *Science (80).* **2019**, *364*, 468.

[17]   P. Luo, F. Zhuge, Q. Zhang, Y. Chen, L. Lv, Y. Huang, H. Li, T. Zhai, *Nanoscle horinos* **2019**, *4*, 26.

[18]   A. Kozhakhmetov, B. Schuler, A. M. Z. Tan, K. A. Cochrane, J. R. Nasr, H. El-Sherif, A. Bansal, A. Vera, V. Bojan, J. M. Redwing, N. Bassim, S. Das, R. G. Hennig, A. Weber-Bargioni, J. A. Robinson, *Adv. Mater.* **2020**, *32*.

[19]   Y. C. Lin, R. Torsi, D. B. Geohegan, J. A. Robinson, K. Xiao, *Adv. Sci.* **2021**, *2004249*, 1.

[20]   M. Li, B. Cai, R. Tian, X. Yu, M. B. H. Breese, X. Chu, Z. Han, S. Li, R. Joshi, A. Vinu, T. Wan, Z. Ao, J. Yi, D. Chu, *Chem. Eng. J.* **2021**, *409*.

[21]   L. Zhang, G. Wang, Y. Zhang, Z. Cao, Y. Wang, T. Cao, C. Wang, B. Cheng, W. Zhang, X. Wan, J. Lin, S. J. Liang, F. Miao, *ACS Nano* **2020**, *14*, 10265.

[22]   J. Zou, Z. Cai, Y. Lai, J. Tan, R. Zhang, S. Feng, G. Wang, J. Lin, B. Liu, H. M. Cheng, *ACS Nano* **2021**, *15*, 7340.

[23]   J. Jiang, L. A. T. Nguyen, T. D. Nguyen, D. H. Luong, D. Y. Kim, Y. Jin, P. Kim, D. L. Duong, Y. H. Lee, *Phys. Rev. B* **2021**, *103*, 014441.

[24]   S. Li, J. Hong, B. Gao, Y. C. Lin, H. E. Lim, X. Lu, J. Wu, S. Liu, Y. Tateyama, Y. Sakuma, K. Tsukagoshi, K. Suenaga, T. Taniguchi, *Adv. Sci.* **2021**, *8*, 2004438.

[25]   S. J. Yun, D. L. Duong, D. M. Ha, K. Singh, T. L. Phan, W. Choi, Y. M. Kim, Y. H. Lee, *Adv. Sci.* **2020**, *7*, 1903076.

[26]   H. Gao, J. Suh, M. C. Cao, A. Y. Joe, F. Mujid, K. H. Lee, S. Xie, P. Poddar, J. U. Lee, K. Kang, P. Kim, D. A. Muller, J. Park, *Nano Lett.* **2020**, *20*, 4095.

[27]   L. M. Dyagileva, V. P. Mar'in, E. I. Tsyganova, G. A. Razuvaev, *J. Organomet. Chem.* **1979**, *175*, 63.

[28]   J. R. Shallenberger, *Surf. Sci. Spectra* **2018**, *25*.

[29]   M. Hossain, J. Wu, W. Wen, H. Liu, X. Wang, L. Xie, *Adv. Mater. Interfaces* **2018**, *5*, 1.

[30]   G. Hopfengärtner, D. Borgmann, I. Rademacher, G. Wedler, E. Hums, G. W. Spitznagel, *J. Electron Spectros. Relat. Phenomena* **1993**, *63*, 91.

[31]   N. D. Boscher, C. S. Blackman, C. J. Carmalt, I. P. Parkin, A. G. Prieto, *Appl. Surf. Sci.* **2007**, *253*, 6041.

[32]   W. Zhao, Z. Ghorannevis, K. K. Amara, J. R. Pang, M. Toh, X. Zhang, C. Kloc, P. H. Tan, G. and

Eda, *Nanoscale* **2013**, *5*, 9677.

[33] E. Del Corro, H. Terrones, A. Elias, C. Fantini, S. Feng, M. A. Nguyen, T. E. Mallouk, M. Terrones, M. A. Pimenta, *ACS Nano* **2014**, *8*, 9629.

[34] W. Shi, M. L. Lin, Q. H. Tan, X. F. Qiao, J. Zhang, P. H. Tan, *2D Mater.* **2016**, *3*.

[35] Q. Qian, Z. Zhang, K. J. Chen, *Langmuir* **2018**, *34*, 2882.

[36] S. Mouri, Y. Miyauchi, K. Matsuda, *Nano Lett.* **2013**, *13*, 5944.

[37] S. J. Yun, Y. Kim, S. Kim, Y. H. Lee, *Appl. Phys. Lett.* **2019**, *115*, 242406.

[38] K. F. Mak, K. He, C. Lee, G. H. Lee, J. Hone, T. F. Heinz, J. Shan, *Nat. Mater.* **2013**, *12*, 207.

[39] B. Schuler, J. H. Lee, C. Kastl, K. A. Cochrane, C. T. Chen, S. Refaely-Abramson, S. Yuan, E. Van Veen, R. Roldán, N. J. Borys, R. J. Koch, S. Aloni, A. M. Schwartzberg, D. F. Ogletree, J. B. Neaton, A. Weber-Bargioni, *ACS Nano* **2019**, *13*, 10520.

[40] S. Barja, S. Refaely-Abramson, B. Schuler, D. Y. Qiu, A. Pulkin, S. Wickenburg, H. Ryu, M. M. Ugeda, C. Kastl, C. Chen, C. Hwang, A. Schwartzberg, S. Aloni, S. K. Mo, D. Frank Ogletree, M. F. Crommie, O. V. Yazyev, S. G. Louie, J. B. Neaton, A. Weber-Bargioni, *Nat. Commun.* **2019**, *10*.

[41] P. Mallet, F. Chiapello, H. Okuno, H. Boukari, M. Jamet, J. Y. Veuillen, *Phys. Rev. Lett.* **2020**, *125*, 036802.

[42] L. Gross, F. Mohn, P. Liljeroth, J. Repp, F. J. Giessibl, G. Meyer, *Science (80).* **2009**, *324*, 1428.

[43] M. Aghajanian, B. Schuler, K. A. Cochrane, J. H. Lee, C. Kastl, J. B. Neaton, A. Weber-Bargioni, A. A. Mostofi, J. Lischner, *Phys. Rev. B* **2020**, *101*.

[44] O. L. Krivanek, M. F. Chisholm, V. Nicolosi, T. J. Pennycook, G. J. Corbin, N. Dellby, M. F. Murfitt, C. S. Own, Z. S. Szilagyi, M. P. Oxley, S. T. Pantelides, S. J. Pennycook, *Nature* **2010**, *464*, 571.

[45] N. Nayir, Y. Wang, S. Shabnam, D. Reifsnyder Hickey, L. Miao, X. Zhang, S. Bachu, N. Alem, J. Redwing, V. H. Crespi, A. C. T. van Duin, *J. Phys. Chem. C* **2020**, *124*, 28285.

[46] S. Wang, A. Robertson, J. H. Warner, *Chem. Soc. Rev* **2018**, *47*, 6764.

[47] D. Jayachandran, A. Oberoi, A. Sebastian, T. H. Choudhury, B. Shankar, J. M. Redwing, S. Das, *Nat. Electron.* **2020**, *3*, 646.

[48] A. Sebastian, R. Pendurthi, T. H. Choudhury, J. M. Redwing, S. Das, *Nat. Commun.* **2021**, *12*.

[49] S. Fan, S. J. Yun, W. J. Yu, Y. H. Lee, *Adv. Sci.* **2019**, 1902751.

[50] F. Mohn, B. Schuler, L. Gross, G. Meyer, *Appl. Phys. Lett.* **2013**, *102*, 1.

[51] K.-K. Liu, W. Zhang, Y.-H. Lee, Y.-C. Lin, M.-T. Chang, C.-Y. Su, C.-S. Chang, H. Li, Y. Shi, H. Zhang, C.-S. Lai, L.-J. Li, *Nano Lett.* **2012**, *12*, 1538.

[52] G. Kresse, J. Furthmüller, *Phys. Rev. B* **1996**, *54*, 11169.

[53] P. E. Blöchl, *Phys. Rev. B* **1994**, *50*, 17953.

[54] G. Kresse, D. Joubert, *Phys. Rev. B* **1999**, *59*, 1758.

[55] J. P. Perdew, K. Burke, M. Ernzerhof, *Phys. Rev. Lett.* **1996**, *77*, 3865.

[56] M. Methfessel, A. T. Paxton, *Phys. Rev. B* **1989**, *40*, 3616.

[57] H. J. Monkhorst, J. D. Pack, *Phys. Rev. B* **1976**, *13*, 5188.

Supporting information

Controllable p-type Doping of 2D WSe$_2$ via Vanadium Substitution

Azimkhan Kozhakhmetov,[1] Samuel Stolz,[2] Anne Marie Z. Tan,[3,4] Rahul Pendurthi,[5] Saiphaneendra Bachu,[1] Furkan Turker,[1] Nasim Alem,[1,6,7] Jessica Kachian,[8] Saptarshi Das,[1,5,6] Richard G. Hennig,[4] Oliver Gröning,[2] Bruno Schuler,[2] Joshua A. Robinson.[1,7,9,*]

1. Department of Materials Science and Engineering, The Pennsylvania State University, University Park, PA 16802, USA
2. nanotech@surfaces Laboratory, Empa – Swiss Federal Laboratories for Materials Science and Technology, 8600 Dübendorf, Switzerland
3. School of Mechanical and Aerospace Engineering, Nanyang Technological University, Singapore, 639798, Singapore
4. Department of Materials Science and Engineering, University of Florida, Gainesville FL 32611, USA
5. Department of Engineering Science and Mechanics, The Pennsylvania State University, University Park, PA 16802, USA
6. Materials Research Institute, The Pennsylvania State University, University Park, PA 16802, USA
7. Two-Dimensional Crystal Consortium, The Pennsylvania State University, University Park, PA 16802, USA
8. Intel Corporation, 2200 Mission College Blvd, Santa Clara, CA 95054, USA
9. Center for 2-Dimensional and Layered Materials, The Pennsylvania State University, University Park, PA 16802, USA

* jrobinson@psu.edu

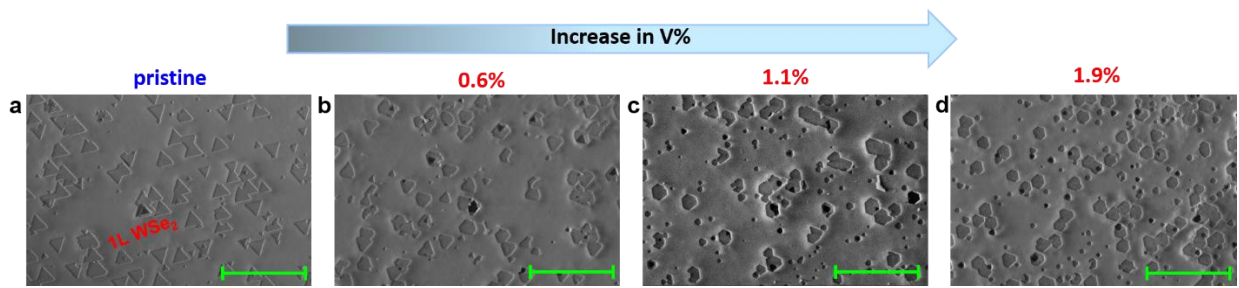

**Figure S1.** Scanning electron microscopy images of (a) pristine and (b) 0.6%, (c) 1.1%, and (d) 1.9% V-WSe$_2$ on sapphire substrates (scale bar 5 μm) demonstrating fully coalesced monolayer film with some secondary islands on the top. Increase in density of secondary nucleation is observed in extrinsic films with the higher vanadium concentration (1.1% and 1.9%).

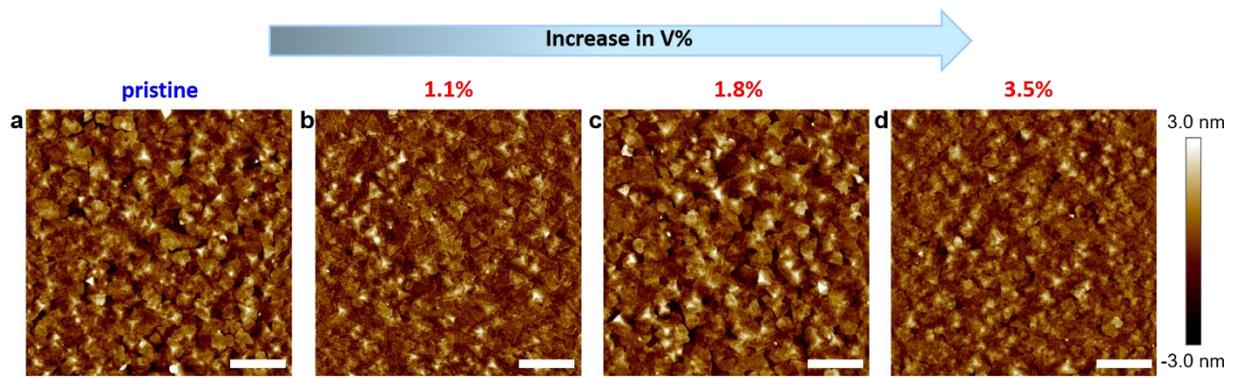

**Figure S2.** AFM images of BEOL compatible (a) intrinsic and (b-d) extrinsic WSe$_2$ films obtained on SiO$_2$/Si substrates at 400 °C (scale bar 1 μm). All obtained films are polycrystalline with an average domain size of 100 nm.

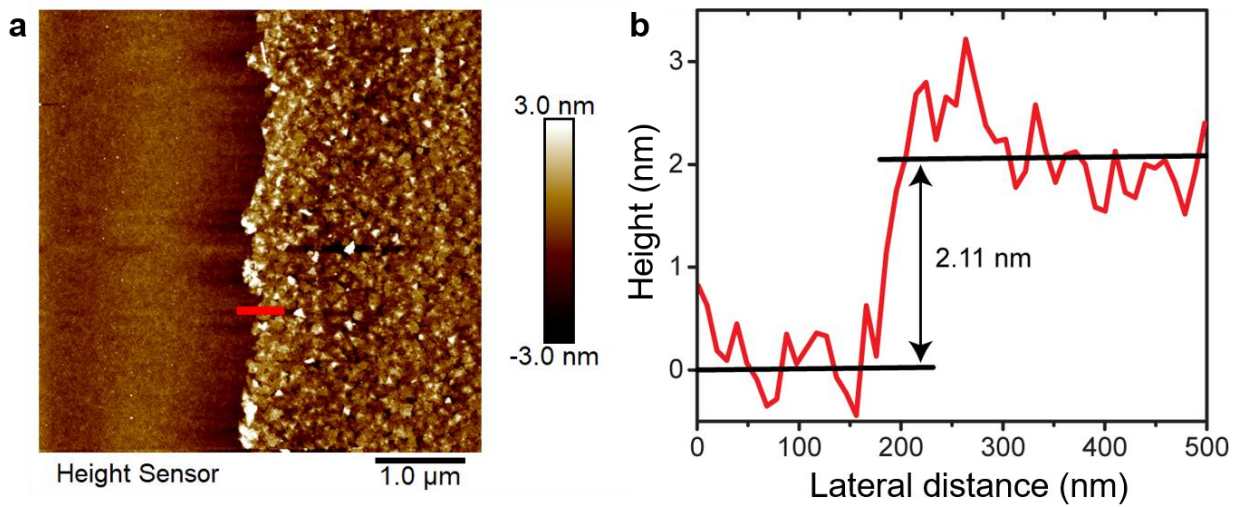

**Figure S3.** AFM image of the surface after a scratch is made and corresponding height profile image of the BEOL 1.8 % V-WSe$_2$ on SiO$_2$/Si demonstrating thickness of 2.11 nm (3-4 layers).

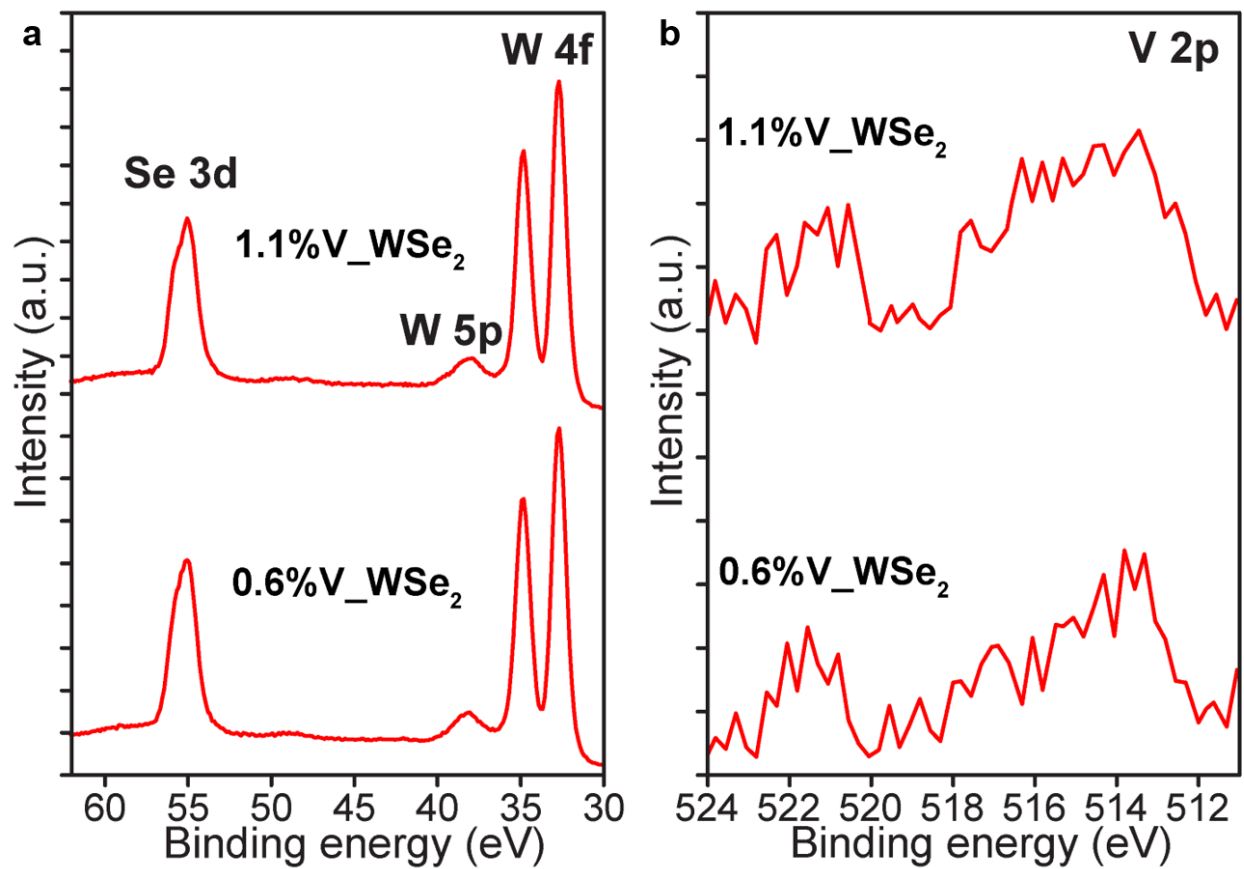

**Figure S4.** High-resolution XPS core level spectra of (a) W 4f, 5p, Se 3d and (b) V 2p regions of FEOL 1.1% and 0.6 % V-WSe$_2$ doped films on sapphire verifying the presence of vanadium atoms in the WSe$_2$ lattice.

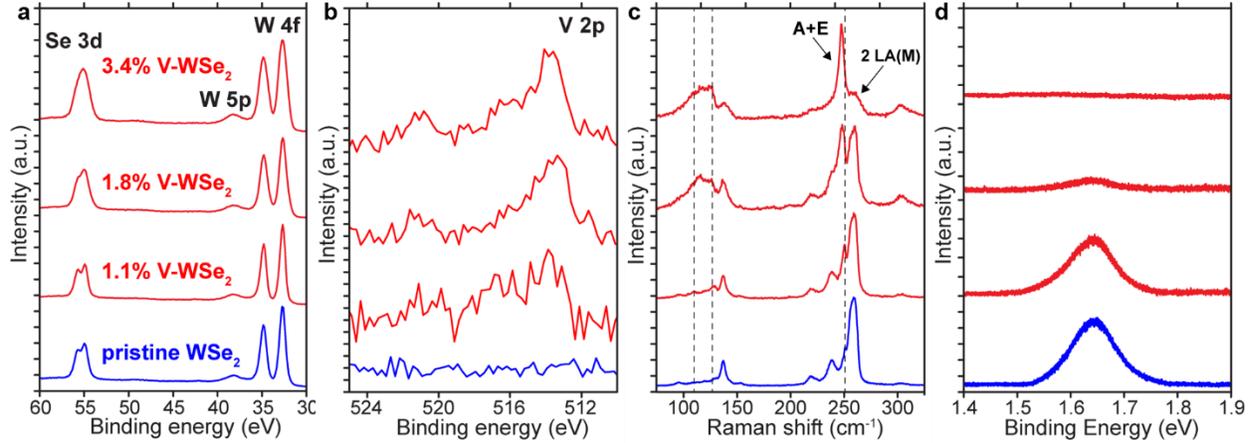

**Figure S5.** Chemical compositional and optical characterization of pristine and doped BEOL WSe$_2$ on SiO$_2$/Si substrates. Core level XPS spectra of (a) W 4f, W5p, Se 3d, and (b) V 2p regions further confirming the presence of vanadium atoms in BEOL films. Corresponding (c) Raman and (b) PL spectra that show similar features as FEOL films on sapphire where A+E mode and defect activated ZA(M), and LA(M) modes get more prominent as the V concentration increases. PL spectra show an optical band gap of 1.62 eV and slowly drops in intensity without redshifting and become almost quenched at concentrations exceeding 1.8%. Note that the A+E mode is redshifted and ZA(M) and LA(M) modes are merged together in BEOL V-WSe$_2$. The aforementioned observations in BEOL V-WSe$_2$ differs from FEOL films and attributed to the difference in crystal quality in which BEOL films are polycrystalline and multilayer with non-uniform thickness distribution whereas FEOL films are mostly monolayer and epitaxial to the underlying sapphire substrate.

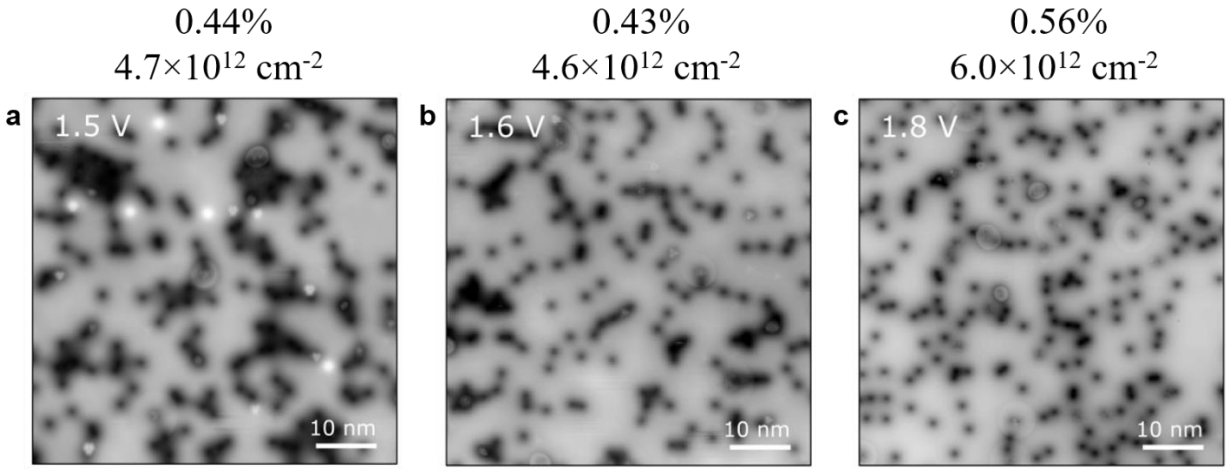

**Figure S6.** STM overview images of three different samples of V-doped WSe$_2$ demonstrating uniform doping distribution on EG surface.

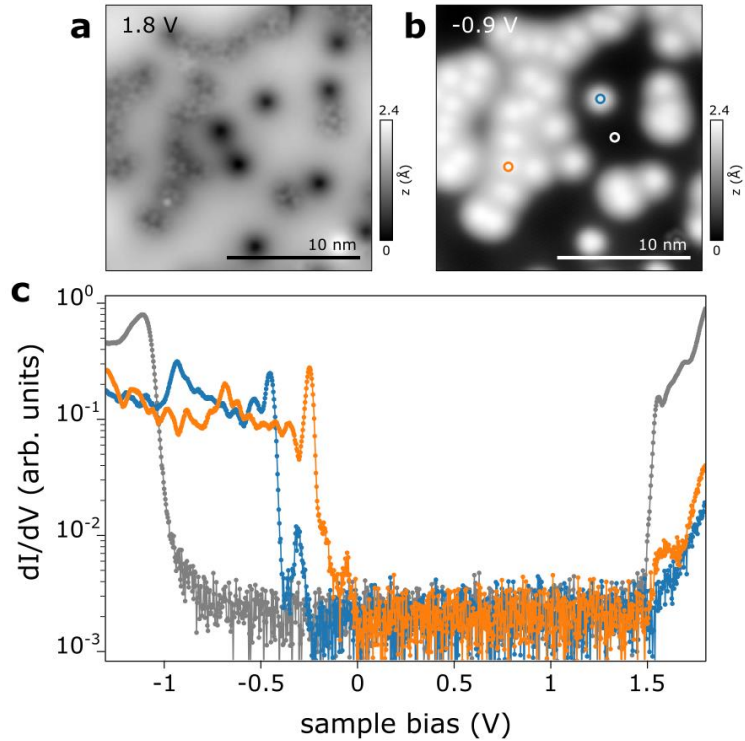

**Figure S7.** (a,b) STM topographies of a densely populated area of V dopants. The dopant's STM contrast changes from a spherical (dark) depression to a dim orbital-like protrusion superimposed on the dark depression at close dopant separations. (c) dI/dV spectra comparing a relatively isolated $V_W$ defect (blue) to a $V_W$ defect in a densely V-doped area (orange). The rigid shift towards higher energies results from an electrostatic interaction with neighboring negatively charged V dopants. Spectra positions are indicated by circles in (b).

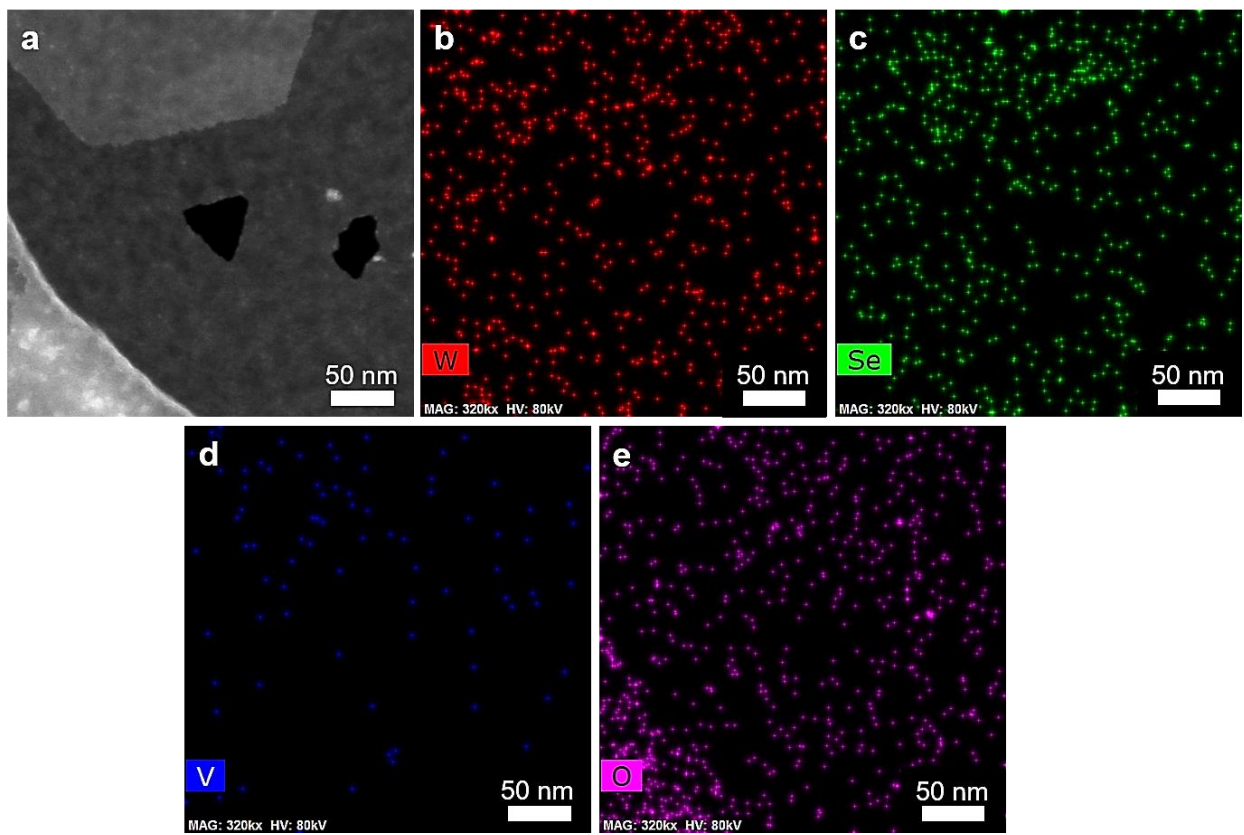

**Figure S8.** (a) HAADF-STEM image of ~ 1.9% V doped WSe$_2$ film suspended over Quantifoil amorphous carbon layer in the TEM grid. (b-e) STEM-EDS maps showing uniform distribution of W, Se, V and O.

**Table S1.** Formation energies ($E^f$) of the neutral V dopant at either of the substitutional sites in the WSe$_2$ lattice, computed by DFT. The reported formation energies are referenced relative to the bcc V chemical potential for $\mu_V$, bcc W for $\mu_W$ (W-rich limit), or hexagonal (gray) Se for $\mu_{Se}$ (Se-rich limit). The substitutional V on W site is found to be the most energetically favorable under both W-rich and Se-rich conditions.

|  | $E^f$ (eV) | |
| --- | --- | --- |
|  | PBE | PBE + SOC |
| $V_W$ | W-rich: -0.12 / Se-rich: -1.78 | -0.33 / -1.93 |
| $V_{Se}$ | 3.14 / 3.97 | 3.12 / 3.92 |

**Table S2.** Key electrical properties of the WSe$_2$ BGFETs for the various doping concentrations. Enhancement in $I_{ON}$ for the p channel and increasing threshold voltage confirm p doping nature of V atoms in WSe$_2$.

|  | Pristine WSe$_2$ | 0.6 % V-WSe$_2$ | 1.1% V-WSe$_2$ | 1.9% V- WSe$_2$ |
|---|---|---|---|---|
| $V_{th}$ (V) | -7.72 | -7.91 | -7.10 | -4.90 |
| $I_{ON}/I_{OFF}$ | 1.32E+06 | 3.05E+06 | 3.17E+06 | 1.65E+06 |
| $\mu_h$ (cm$^2$/V-s) | 3.00 | 1.75 | 0.83 | 1.91 |
| $SS_h$ (mV/dec) | 406 | 553 | 747 | 976 |
| $I_{ON}$ (µA/µm) | 0.96 | 0.92 | 1.85 | 2.89 |

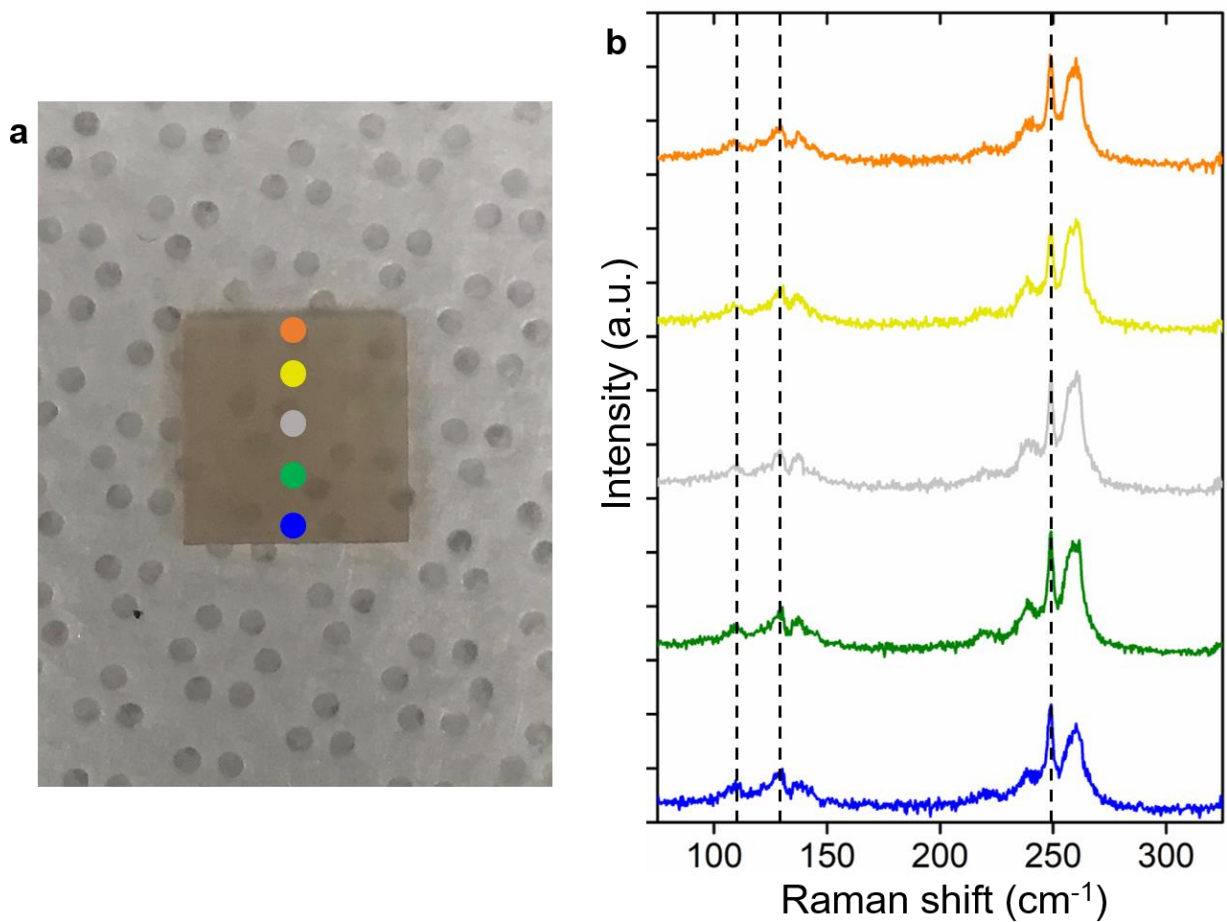

**Figure S9.** (a) Optical image of FEOL 1.1% V-WSe$_2$ on c-plane sapphire substrates (substrate size is 1 cm × 1 cm) and (b) corresponding Raman spectra obtained from the entire substrate surface. The increased intensity of ZA(M), LA(M), and A+E modes at 109.2 cm$^{-1}$, 128.6 cm$^{-1}$, and 249.3 cm$^{-1}$, respectively, (dashed lines) confirms that the dopant atoms are uniformly distributed all over the substrate.

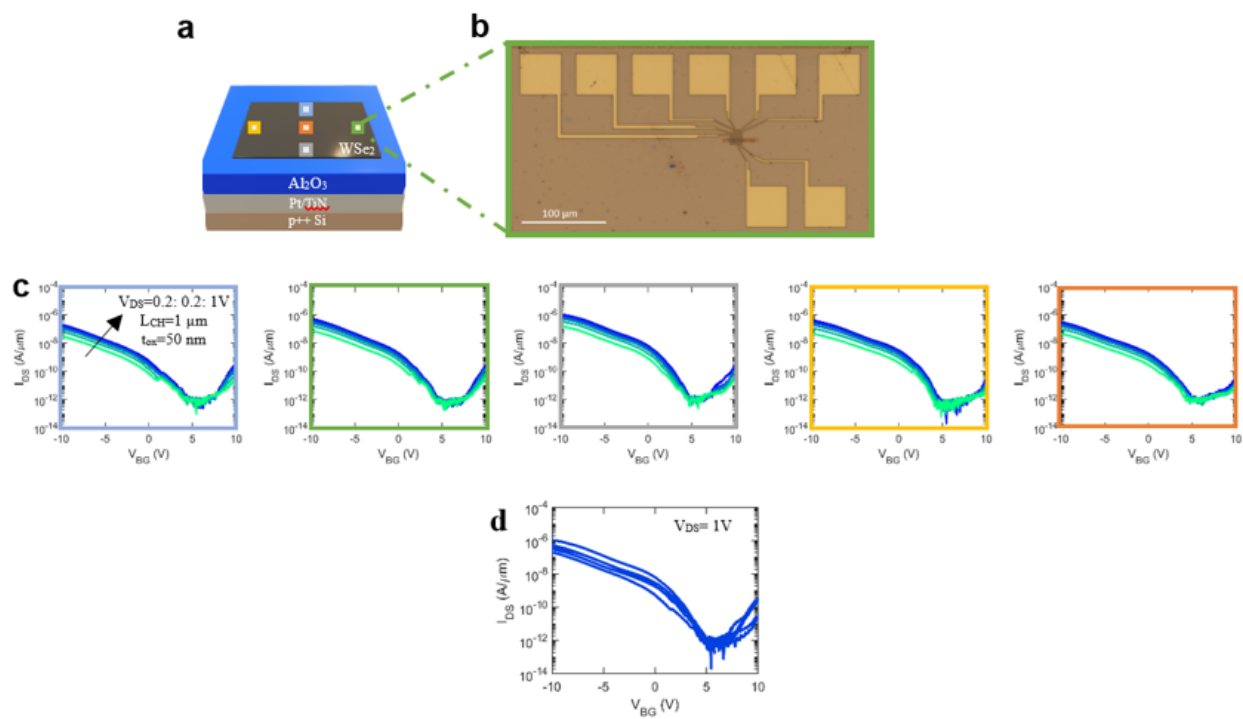

**Figure S10.** (a) Representative schematic of the transferred 1.1% V-WSe$_2$ film on the 50nm Al$_2$O$_3$ substrate. Five devices are measured from different regions of the film, with an (b) optical image of the 1µm channel length BGFET device structure. (c) Transfer characteristics of the BGFETs selected at different regions on the film. (d) Device variability across the film for 5 devices at $V_{DS}$=1V.